\documentclass{llncs}
\usepackage{amsmath,amssymb,amsfonts}
\usepackage{footnote}
\usepackage{graphicx}
\usepackage{pdflscape} 
\usepackage{hyperref}
\usepackage[braket, qm]{qcircuit}
\usepackage{mathtools}
\usepackage[font=small, labelfont={sf,bf}, margin=1cm]{caption}
\usepackage{caption}
\usepackage{subcaption}
\usepackage{color}
\usepackage{pdfpages}
\usepackage{listings}
\usepackage{comment}
\usepackage{xcolor}

\DeclarePairedDelimiter\ceil{\lceil}{\rceil}
\DeclarePairedDelimiter\floor{\lfloor}{\rfloor}

\newcommand{\tf}{\textit{Toffoli} }
\newcommand{\cx}{\textit{CNOT} }
\newcommand{\sio}{$ SIMON $ }

\begin{document}
	
\title{Grover on \sio}

\author{Ravi Anand$^1$, Arpita Maitra$^{2,3}$, Sourav Mukhopadhyay$^1$}
\institute{$^1$ Department of Mathematics, Indian Institute of Technology Kharagpur,  Kharagpur-721302, West Bengal, India\\
$^2$ TCG Centre for Research and Education in Science and Technology, Kolkata-700091, West Bengal, India. \\
 $^3$CR Rao Advanced Institute of Mathematics, Statistics and Computer Science, Hyderabad, India.}
\maketitle

	\begin{abstract}
 For any symmetric key cryptosystem with $n$-bit secret key, 
the key can be recovered in $O(2^{n/2})$ exploiting  Grover search algorithm,  resulting in the effective key length 
to be half.  In this direction, subsequent work has been done on AES and some other block ciphers. On the other hand, lightweight ciphers like \sio was left unexplored. In this backdrop, we present  Grover's search algorithm on all the variants of \sio and  enumerate the quantum resources to implement such attack in terms of NOT, CNOT and Toffoli gates. We also provide the T-depth of the circuits and the number of qubits required for the attack. We show that the number of qubits required for implementing Grover on  \sio $2n/mn$ is $O(2nr+mn)$, where $r$ is the number of chosen plaintext-cipher text pairs. 
We run a reduced version of \sio in IBMQ quantum simulator and the 14-qubits processor as well. We found that where simulation supports theory, the actual implementation is far from the reality due to the infidelity of the gates and short decoherence time of the qubits. The complete codes for all version of \sio have also been presented.
	\end{abstract}
	
	\section{Introduction}
	The last two decades witnessed an enormous proliferation in the domain of quantum computation and communication. Due to the two pioneering quantum algorithms, Shor's algorithm~\cite{shor1999polynomial} and Grover's search algorithm~\cite{grover}, the security of currently deployed cryptosystems is under a threat. As a consequence, in recent time, a lot of symmetric constructions are being evaluated in quantum settings. For example, one can mention the key recovery attacks against Even-Mansour constructions and distinguishers against 3-round Feistel constructions~\cite{3roun}. Not only that, the key recovery attacks against multiple encryptions~\cite{kaplan}, forgery attacks against CBC-like MACs~\cite{kaplan1} have also been studied.  The list is expanding considering Quantum meet-in-the-middle attack on Feistel constructions~\cite{demeric}, key recovery attacks against FX constructions~\cite{fx} etc. Researchers have also tried to convert the existing classical attacks to quantum settings~\cite{kaplan2,hs,kaplan1,tho}.
	
	Very recently, Bonnetain et al.~\cite{bonnetain} proposed a novel methodology for quantum cryptanalysis on symmetric ciphers. They exploited the algebraic structure of certain classical cryptosystems to bypass the standard quantum superposition queries.  
	
	 In case of symmetric ciphers or hash function, Grover's algorithm provides a quadratic speed up in exhaustive key search. So a conservative rule of thumb is to double the security parameter, i.e.,  atleast double the size of the key or double the size of the output of hash function. However, this does not rule out the need of  analyzing the cost of Grover's algorithm on symmetric ciphers. In this direction, subsequent efforts have been made to derive cost estimation for applying Grover's search algorithm on all variants of AES~\cite{aes,aes1,aes2,aes3}. The cost of applying Grover's search algorithm as a pre-image search attack on hash functions has also been studied~\cite{hash}.
	
	 Fault tolerant commercialized quantum computers are still elusive. However, several companies are providing simulation facilities through the web. Along with the simulation, IBM provides facility to run the program in their small scale actual quantum processors. Based on this state-of-the-art situation, this is very important to explore actual implementation issues of all these quantum cryptanalysis procedures.

		\sio is a family of lightweight block ciphers released by NSA in June $ 2013 $. It is a balanced Feistel structured block cipher. \sio is optimized for performance in hardware implementations. In October $ 2018 $, the \sio and Speck ciphers have been standardized by ISO as a part of the RFID Air Interface Standard, International Standard ISO/29167-21 (for \sio) and International Standard ISO/29167-22 (for Speck), making them available for use by commercial entities~\cite{wiki} 
	
	As these are comparatively new ciphers, quantum cryptanalysis on those ciphers remained unexplored. In this backdrop, in the current effort, we study the cost of Grover search on all the variants of \sio and try to implement that in publicly available IBM quantum processors. Due to the limitation of the qubits, we could not run the full scale cipher, instead we run the algorithm for a reduced version. We found that whereas simulation meets theory, actual implementation has been masked with error. \\
	
	\noindent \textbf{Our Contribution.} One may argue that it is already well known that Grover search provides quadratic speedup over classical exhaustive key search. In this direction, we like to emphasize that for implementing Grover algorithm on a symmetric cipher, one requires a reversible implementation of that cipher which is a hard task. In this regard, we design the reversible version of all the variants of \sio~\cite{\sio} so that one can successfully implement Grover oracle and Grover diffusion for key search on all of those variants. We also provide the full implementation code for the cipher in QISKIT~\cite{qiskit}. We estimate the resources in terms of NOT, CNOT, and Toffoli gates required to attack the cipher. We provide the T-depth of the circuits and the number of qubits needed to implement the attack, too. We tested our circuits with existing classical test vectors to make sure that the implementations are correct.   
	The code for all the variants is given in~\cite{grover1}. This is because when the full scale quantum computers arrive, one may implement the code immediately. For independent verification of our results with the state-of-the-art IBM quantum simulator and processors, we add all the QASM and QISKIT codes for a reduced \sio. To the best of our knowledge, this is the first full implementation and resource estimation on \sio in quantum settings.

	\section{Preliminaries}
	\label{preliminaries}

	\subsection{Brief Summary on \sio}
	\label{description_of_\sio}
\begin{figure}[ht]
		\centering
		\includegraphics[scale = 0.6]{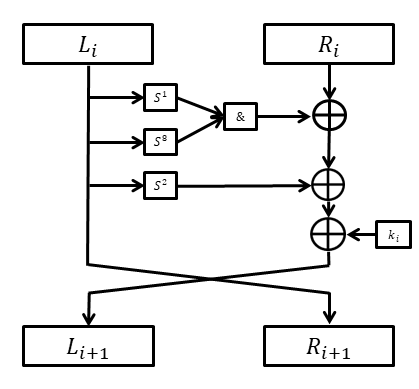}
		\caption{\sio round function}
		\label{sim}
	\end{figure}

	\sio is a family of balanced Feistel structured lightweight block ciphers with 10 different block sizes and key sizes (Table~\ref{Spara}). The round function used in the Feistel structure of \sio block ciphers consists of circular shift, bitwise AND and bitwise XOR operations. The state update function is defined as,
	\begin{align}
	F(x,y) = (y \oplus S^1 (x) S^8 (x) \oplus S^2 (x) \oplus k, x)
	\label{supdate}
	\end{align}
	The structure of one round \sio encryption is depicted in Figure~\ref{sim}, where $ S^j $ represents a left circular shift by $ j$ bits, $L_i$ and $ R_i $ are $n$-bit words which constitutes the state of \sio at the $i$-th round and $ k_i $ is the round key which is generated by key scheduling algorithm described below.\\ 
	
	The different variants of \sio are denoted by \sio$2n/mn$, where $2n$ denotes the block size of the variant, and $mn$ is the size of the secret key. 
	Here $n$ can take values from $16, 24, 32, 48$ or $64$, and $m$ from $2, 3$ or $4$. For each combination of $(m, n)$, the corresponding round number $T$ is 
	adopted.
	\begin{table}[ht]
		\centering
		\scalebox{1.0}{
			\begin{tabular}{|c|c|c|c|c|}
				\hline
				Block Size $(2n)$ & Key Size $(k=mn)$ & word size $(n)$  & keywords $(m)$ & Rounds $(T)$ \\
				\hline
				32 & 64 & 16 & 4 & 32\\
				\hline
				48 & 72,96 & 24 & 3,4 & 36,36\\
				\hline
				64 & 96,128 & 32 & 3,4 & 42,44\\
				\hline
				96 & 96, 144 & 48 & 2,3 & 52,54\\
				\hline
				128 & 128,192,256 & 64 & 2,3,4 & 68,69,72\\
				\hline
		\end{tabular}}
		\caption{\sio parameters}
		\label{Spara}
	\end{table}

	The key schedule  of \sio has three different procedures depending on the key size. The first $m$ round keys are initialized directly from the main key. The remaining $(T-m)$ round keys are generated by the following procedure:\\
	\begin{align}
	k_{i+m} = \begin{cases}
	c_i \oplus  k_i  \oplus  S^{-3}(k_{i+1})  \oplus S^{-4}(k_{i+1}) & m=2,\\
	c_i \oplus  k_i  \oplus  S^{-3}(k_{i+2})  \oplus S^{-4}(k_{i+2}) & m=3,\\
	c_i \oplus  k_i  \oplus  S^{-1}(k_{i+1})  \oplus  S^{-3}(k_{i+3})  \oplus S^{-4}(k_{i+3}) & m=4,\\
	\end{cases}
	\label{keyexp}
	\end{align} \\
	where $c_i$ are the round dependent constants and $S^{-j}(x)$ denotes right rotation by $j$ times on $x$. 
		
\subsubsection{Encryption: }
	The input to the encryption oracle is a $ 2n $-bit plaintext block $ P $. This block is divided into $ n- $bit subblocks $P =  (L_0, R_0) $ which is the initial state of the cipher. The encryption consists of $ T $ applications of the round function with the respective round key produced by the key schedule. The ciphertext obtained is a $ 2n- $bit block $ C = (L_{T-1},R_{T-1}) $
	
	\subsubsection{Decryption:}
	Decryption of the ciphertext $ C = (L_{T-1},R_{T-1}) $ consists of first swapping $ L  $ and $ R $ part of the block cipher, i.e. the input to the decryption oracle is $ (R_{T-1},L_{T-1}) $. Then  $ T $ round functions with round
	keys in reverse order (ie. round keys $ k_{T-1},\cdots k_0 $) is applied followed by a final swapping of the two subblocks. \\

	\subsubsection{Existing cryptanalysis of SIMON:} 
	To the best of our knowledge, no successful attack on full-round Simon of any variant is known. As is typical for iterated ciphers, reduced-round variants have been successfully attacked. Some of the results are summarized below Table~\ref{sum}.
	\begin{table}[h]
		\centering
		\scalebox{0.85}{
		\begin{tabular}{|c|c|c|c|c|}
			\hline
			Variant	Rounds& attacked&	Time complexity&	Data complexity&	Attack type\\
			\hline
			Simon32/64&	21/32 &	$ 2^{63} $&	$ 2^{31} $&	Integral~\cite{wang}\\
			\hline
			Simon48/72&	20/36 &	$ 2^{59.7} $&	$ 2^{48} $&Zero Correlation~\cite{wang} \\
			\hline
			Simon48/96&	21/36 &	$ 2^{72.63} $	&$ 2^{48} $&	Zero Correlation~\cite{wang}\\
			\hline
			Simon64/96&	26/42 &	$ 2^{63.9} $&	$ 2^{63} $&	Differential~\cite{abed}\\
			\hline
			Simon64/128&	26/44 &	$ 2^{94} $&	$ 2^{63} $&	Differential~\cite{abed}\\
			\hline
			Simon96/96&	35/52 &	$ 2^{93.3} $&$ 	2^{93.2} $&	Differential~\cite{abed}\\
			\hline
			Simon96/144	&35/54& 	$ 2^{101} $&	$ 2^{93.2} $&	Differential~\cite{abed}\\
			\hline
			Simon128/128&	46/68& 	$ 2^{125.7} $&$ 	2^{125.6} $&Differential~\cite{abed}\\
			\hline
			Simon128/192&	46/69 &	$ 2^{142} $&	$ 2^{125.6} $&	Differential~\cite{abed}\\
			\hline
			Simon128/256 &	46/72 &	$ 2^{206} $&	$ 2^{125.6} $&	Differential~\cite{abed}\\
			\hline	
		\end{tabular}}
	\caption{Summary of existing cryptanalysis results on \sio}
	\label{sum}
	\end{table}
	For a more detailed description of \sio, the readers are referred to \cite{\sio}.\\
	
	\subsection{Grover's Algorithm}
	\label{description of Grover's Algorithm}
	Grover's algorithm \cite{grover} searches through a space of $  N $ elements for a solution. We can assume that $N=2^n$ and each state be represented by the indices in $\{0,1\}^n$. Let us assume that there exists a state $y$ such that\\
	\begin{align*}
	f(x) = \begin{cases}
	1 & \text{if }$ x=y $,\\
	0 & \text{otherwise}.\\
	\end{cases}
	\end{align*}
	We also assume that $f$ is easily and effectively computable. The \textit{oracle}, $f$ is provided as a black box. It finds the state $y$. Grover's algorithm needs only $O(2^{n/2})$ oracle calls as compared to $O(2^n)$ oracle calls needed classically.\\
	
\noindent  Grover's algorithm can be summarised in the following steps:
	\begin{enumerate}
		\item Apply Hadamard gate on the initial state $\ket{00...0}$ bit by bit to obtain the following superposition $$ \ket{\psi} = \frac{1}{2^{n/2}} \sum\limits_{x=0}^{2^{n}-1} \ket{x}$$
		\item The second step make $\floor{\frac{\pi}{4}2^{n/2}}$ calls to  Grover's iteration.  Grover's iteration comprises of two subroutines.\\
		The first subroutine makes use of the operator $U_f$ which evaluates the Boolean function $f:\{0,1\}^n \rightarrow \{0,1\}$, which marks the solutions of the search problem, $ i.e.~ f(x) = 1 $ if and only if the element corresponding to $x$ is a solution. When we apply the Gorver oracle $U_f$ to a state $ \ket{x}\ket{z} $, where $ \ket{x} $ is a $n$-qubit state and $\ket{z}$ is a single qubit then it acts as $U_f:\ket{x}\ket{z} \rightarrow \ket{x}\ket{z \oplus f(x)}$. If $\ket{z}$ is chosen to be $\frac{1}{\sqrt{2}}(\ket{0} - \ket{1})$, then we have $U_f:\ket{x}\frac{1}{\sqrt{2}}(\ket{0} - \ket{1}) \rightarrow (-1)^{f(x)}\ket{x}\frac{1}{\sqrt{2}}(\ket{0} - \ket{1})$ which means that the oracle applies a phase shift only to the solution indices while leaving the remaining indices unaltered. Each call to $U_f$ involves two calls to a reversible implementation of $f$ and one call to a comparision circuit that checks if $x$ is a solution or not.\\
		
		The second subroutine implements the transformation $2\ket{0}\bra{0}- I$, also known as the diffusion operator. This routine flips the amplitude of the states about it's mean thus amplifying the amplitude of the solution. This involves single qubit  gates and one $t$-fold controlled $NOT$ gate. So in the second step, the following two steps are repeated $O(2^{n/2})$ times:
		\begin{enumerate}
			\item For any state $\ket{x}$ in the superposition $\ket{\psi}$, rotate the phase by $\pi$ radians if $f(x) = 1$ and leave the system unaltered otherwise.
			\item Apply the diffusion operator. 
		\end{enumerate}
		\item Measure the resulting superposition and obtain the solution with the probabilities determined by the amplitudes of the states. 
	\end{enumerate}
	
	\subsection{Block Cipher Key Search Using Grover}
	Let $E$ be a block cipher with block size $n$ and key size $k$. For any key $K \in \{0,1\}^k$, let $E_K(M)$ be the encryption of plaintext $M$ under the key $K$. For a given plaintext-ciphertext pair $(M,C)$ with $C = E_K(M)$, we can apply Grover's algorithm to determine the key $K$ \cite{qcofblock}. The steps involved are:\\
	\begin{enumerate}
		\item Define a Boolean function $f$ for  Grover's oracle which takes the key $K$ as input,
		\begin{align*}
		f(K) = \begin{cases}
		1 & \text{if } E_{K_0}(M)=C,\\
		0 & \text{otherwise}.
		\end{cases}
		\end{align*}
		\item Initialize the system by making a superposition of all the possible keys with same amplitude,
		\begin{align*}
		\ket{\mathcal{K}} = \frac{1}{2^{K/2}} \sum\limits_{j=0}^{2^{K}-1} \ket{K_j}.
		\end{align*}
		\item Iterate $2(a),(b)$ as described in Section~\ref{description of Grover's Algorithm} for $O(2^{K/2})$ times.
		\item Measure the system and observe the state $K= K_0$ with probability atleast $(\frac{1}{2})$.
	\end{enumerate}

	As a matter of fact, there may be more than one key that satisfies $C=E_{K_0}(M)$. To ensure that the key obtained is unique we may require more than one plaintext-ciphertext pairs under the same key. Let us consider that we have $r$ such pairs ($M_i,C_i$). In this case the Boolean function for  Grover's oracle would be defined as, 
	\begin{align*}
		f(K) = \begin{cases}
		1 & \text{if } E_{K}(M_i) = C_i, 0\leq i \leq r,\\
		0 & \text{otherwise}.
		\end{cases}
	\end{align*}

\subsection{Attack Model}
	We mount known plaintext attack using Grover's algorithm~\cite{grover} for all the variants of \sio . We consider that the adversary has access to certain pairs of plaintexts and corresponding ciphertexts. Then he finds the secret key using Grover's search using quantum resources. In this regard, we need to design a quantum circuit for all the variants of \sio. In the following section, we describe the circuit.\\
	\section{Quantum Circuit for \sio}
	\label{circuits}
	In this section, we develop a reversible quantum circuit for \sio. We analyze our circuits based on the number of qubits, $NOT$ gates, \cx gates, and \tf gates. Grover search will be executed on this circuit under the known plaintext attack model, i.e. when pairs of plaintext and the corresponding ciphertext are already known. 
	
	The circuits described in this section are implemented in QISKIT~\cite{qiskit}. The circuit is reversibly computable and needs no ancilla qubits. We also estimate the T-depth of the circuit.
	
	The internal state size of \sio varies from $32$ bits to $128$ bits as described in Table~\ref{Spara}.
	\sio consists of two subroutines, the round function, and the key expansion. We describe both these routines first and then show how they can be used  simultaneously in the whole cipher construction.
	
	\subsection{Circuit for Round Update Function}
	The round function $F$ is defined as, 
	\begin{align*}
	F(x,y) = (y \oplus S^1 (x) S^8 (x) \oplus S^2 (x) \oplus k, x),
	\end{align*}
	where $S^{i}(x)$ denotes left rotation by $i$ times on $x$.
	
	Now, we assume that we have $k$-qubits reserved for the key, $K$ and, $n$-qubits each for $L$ and $R$. 
	Let $(L_0,R_0)$ be the initial state and the state propagate as $(L_0,R_0),(L_1,R_1),(L_2,R_2), \cdots , (L_j,R_j)$ in $j$ rounds.
	
	\noindent	Now, due to the construction of \sio, we can write
	\begin{align*}
	R_2(i) = L_1(i)  ~= &~R_0(i) ~\oplus ~K_0(i) ~\oplus ~L_0({(i+1)mod(n/2)}) ~\& ~L_0({(i+8)mod(n/2)})\\   &~\oplus ~L_0({(i+2)mod(n/2)}), 0\leq i \leq (n/2),
	\end{align*}
	Note that here $i$ denotes the position of the bit in $L$ and $R$ of a round. If we consider two round \sio, then each bit of $R_2$ will be the XORing of each bit of $R_0$, $F(L_0)$ and $K_0$, where $F(x)=S^1 (x) S^8 (x) \oplus S^2 (x)$. Similarly, each bit of $L_2$ will be the XORing of each bit of $L_0$, $F(R_2)$ and $K_1$. So the qubits reserved for $R_0$ can be used to store the values for $R_2$. Similarly the qubits reserved for $L_0$ can be used to store the value of $L_2$; hence, no need for extra qubits.
	
	Each bit of $R_2$, i.e., $R_2(i)$, is computed using the following three steps: 
	\begin{enumerate}
		\item $ Toffoli(L_0({(i+1)mod(n/2)}),L_0({(i+8)mod(n/2)}),R_0(i)), $
		\item $ CNOT(L_0({(i+2)mod(n/2)}),R_0(i)), $
		\item $ CNOT(K_0(i),R_0(i)). $
	\end{enumerate}
	$L_2$ can be implemented in similar way. Proceeding sequentially, we can build a circuit for as many rounds as required.
	
	Now, it is easy to calculate that for $1$ round we require $n$  \tf gates and $2n$  \cx gates. So, for $j$ rounds we need $jn$  \tf gates and $2jn$  \cx gates.
	
	Let us now define three functions $\mathcal{F},\mathcal{G} \text{ and } \mathcal{H}$, shown below in Figure~\ref{rf}, for easy understanding of the circuit construction.
	
	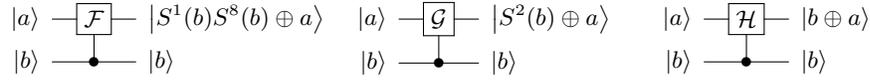
\begin{figure}[ht]	
		\centering
		$\Qcircuit @C=1em @R=1em {
			\lstick{\ket{a}} & \gate{ \mathcal{F} } & \rstick{\ket{S^1(b)S^8(b)\oplus a}} \qw \\
			\lstick{\ket{b}} & \ctrl{-1}  & \rstick{\ket{b}} \qw
		} 	\hspace{3.5cm}
		\Qcircuit @C=1em @R=1em {
			\lstick{\ket{a}} & \gate{\mathcal{G}} & \rstick{\ket{S^2(b)\oplus a}} \qw \\
			\lstick{\ket{b}} & \ctrl{-1}  & \rstick{\ket{b}} \qw
		} \hspace{3cm}
	\Qcircuit @C=1em @R=1em {
	\lstick{\ket{a}} & \gate{\mathcal{H}} & \rstick{\ket{b\oplus a}} \qw \\
	\lstick{\ket{b}} & \ctrl{-1}  & \rstick{\ket{b}} \qw
	} $		
	\caption{Subrotines comprising the round function $F$}
	\label{rf}	
	\end{figure}

	Here $\ket{a} = \ket{a_1a_2...a_n}$, $\ket{b} = \ket{b_1b_2...b_n}$ are $n$ length quantum states and $(+)$ is addition modulo $n$, i.e. $(i+8) = (i+8)mod(n)$.\\
	
\begin{figure}[ht]
		\centering
	$\Qcircuit @C=1.5em @R=1.5em {
						   &                      & \mbox{Round 1}       &                      &     &      &          & &             &  \mbox{Round 2}     &            
		&     &   \\
		\lstick{\ket{K_0}} & \qw                  & \qw \                & \ctrl{1}             & \qw & &       \lstick{\ket{K_1}}  & \qw  & \qw & \qw           & \ctrl{2}  
		& \qw  & & \lstick{\ket{K_2}}  \\
		\lstick{\ket{R_0}}   & \gate{ \mathcal{F} } & \gate{ \mathcal{G} } & \gate{ \mathcal{H} } & \qw &  &  \lstick{\ket{L_1}} & \qw  & \ctrl{1}             & \ctrl{1}             & \qw       
		& \qw &  & \lstick{\ket{R_2}}  \\
		\lstick{\ket{L_0}}   & \ctrl{-1}            & \ctrl{-1}            & \qw                  & \qw & &  \lstick{\ket{R_1}} & \qw  & \gate{ \mathcal{F} } & \gate{ \mathcal{G} } & \gate{\mathcal{H} } & \qw & & \lstick{\ket{L_2}} \\
	}$
	\caption{Circuit for two rounds}
	\label{2rcir}
	\end{figure}
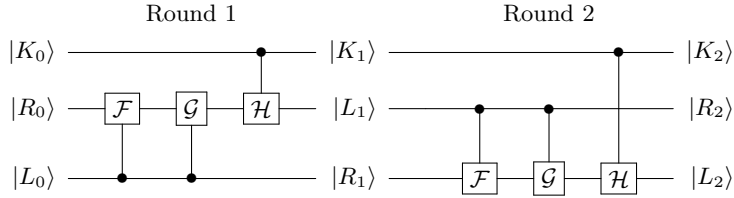

	\noindent The circuit for the two rounds is shown Figure~\ref{2rcir}, where $K_j,L_j,R_j$ represent quantum states of size $n$ for a round $j$.\\

\begin{figure}[ht]
			\includegraphics[scale = 0.15]{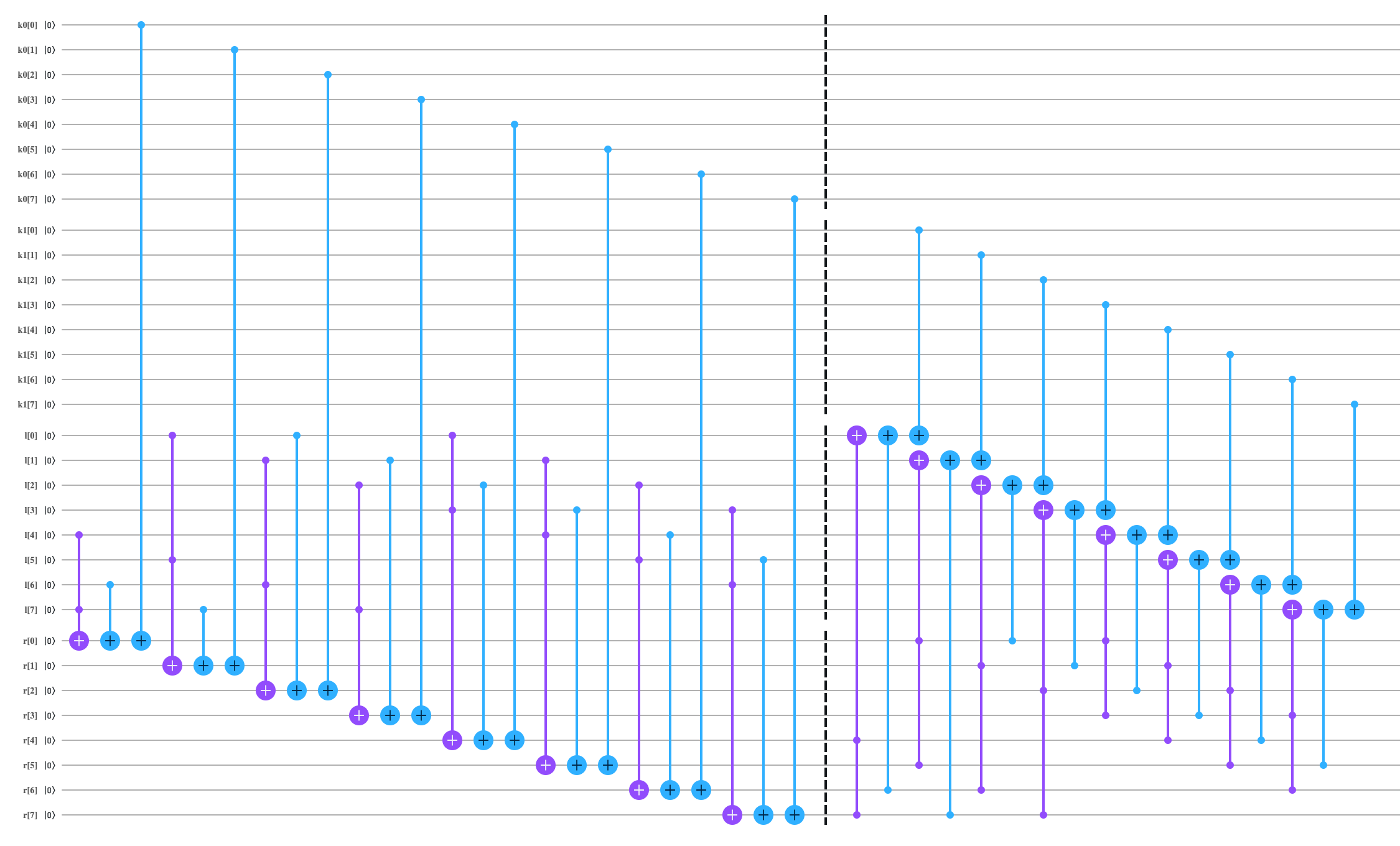}
		\caption{Circuit for two rounds of reduced version of \sio}
		\label{2roundscir}
		\end{figure}

	Now, consider the circuit in Figure~\ref{2roundscir}. This is the IBMQ~\cite{ibmq} implementation of the circuit for two rounds of a reduced version of \sio. The assumed state size is $16$ and the key size is $16$ and $m = 2$. The state is split into $L,R$ each of size $8$ and the key is split into two round keys $(k_0,k_1)$ also of size $8$. The state update function is assumed to be $F(x,y) = (y \oplus S^1 (x) S^4 (x) \oplus S^2 (x) \oplus k, x)$. The circuit here describes the two rounds of the cipher,  i.e.  if we measure the $L$ and $R$ states, we would get the values of the state after two rounds. We can extend these circuits for more than two rounds described earlier.
	
	One should note that this implementation works for all variants of \sio except \sio$128/192$. The problem arises for the \sio$128/192$ as the number of rounds is $69$. This can be solved by implementing the last round such that the state $R$ is modified according to the state update function and $L$ state is left as it is. Then we apply a swap function on the states $L$ and $R$, which increase the number of \cx gates in the circuit by an amount of $64 \times 3 = 192$. 
	
	\subsection{Key Expansion}\label{keyex}
	The key expansion routine is linear and invertible. It is defined in Eqn.~\ref{keyexp}.
	We can implement an in-place construction of key expansion without the use of any ancilla qubits. We here assume that the round constants are implemented in the circuits using an adequate number of $NOT$ gates. This number will depend on the round constant values.
	Here we show the construction for all the three cases $m = 2,3,4$.	
	
	Let us define a subroutine $\mathcal{R}_q$ on two states  $\ket{a} = \ket{a_1a_2...a_n}$ and $\ket{b} = \ket{b_1b_2...b_n}$ such that 
	\begin{align*}
	\mathcal{R}_q(a,b) = (a \oplus S^{-i}(b),b)
	\end{align*}
 That is,  the state $\ket{b}$ remains unchanged and each qubit in $\ket{a}$ gets modified as $\ket{a_j} = \cx ~(b_{(j-i)mod(n)},a_j)$. So each application of $\mathcal{R}_q$ on $\ket{a},\ket{b}$ involves $n$ \cx gates, where $n$ is the size of $\ket{a},\ket{b}$.

	\begin{figure}[ht]
		\centering
		$\Qcircuit @C=1.5em @R=1.5em {
			\lstick{\ket{a}} & \gate{ \mathcal{R}_q } & \rstick{\ket{S^{-i}(b)\oplus a_{i}}} \qw \\
			\lstick{\ket{b}} & \ctrl{-1}  & \rstick{\ket{b}} \qw
		}$
	\end{figure}
	
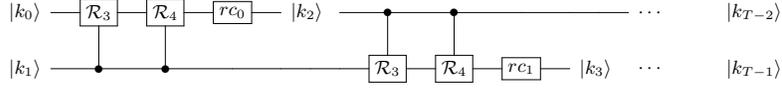
\begin{figure}[ht]
			\centering
			\scalebox{0.8}{
			$\Qcircuit @C=1.5em @R=1.5em {
				\lstick{\ket{k_0}} & \gate{ \mathcal{R}_3} & \gate{ \mathcal{R}_4} & \gate{rc_0}  & \rstick{\ket{k_2}} \qw & & & \ctrl{1} & \ctrl{1} & \qw  &  \qw & \qw&\rstick{\cdots}\qw &  && \rstick{\ket{k_{T-2}}}\\
				\lstick{\ket{k_1}} & \ctrl{-1} & \ctrl{-1} & \qw  & \qw & \qw & \qw &\gate{ \mathcal{R}_3} & \gate{ \mathcal{R}_4} & \gate{rc_1}   & \rstick{\ket{k_3}} \qw &  &\rstick{\cdots} &  &&\rstick{\ket{k_{T-1}}}
			}$}
			\caption{Circuit for key expansion with $2$ keywords. $rc_i$ represents the round dependent constants}.
			\label{w2}
		\end{figure}

	For $m = 2$, we have two key words $k_0,k_1$ which are used for the first and second round of encryption. $k_0,k_1$ are states of size $n$, so we need $2n$ qubits to store this value.\\ The third round key $k_2$ can be computed on the same qubits which will store $k_0$. Each bit $k_2(j)$ can be computed from $(k_0,k_1)$ by applying the following three steps:
	\begin{enumerate}
		\item $ \cx(k_1(j-3)mod(n/2),k_0(j)), $
		\item $ \cx(k_1(j-4)mod(n/2),k_0(j)),$
		\item $ NOT(k_0(j)) \text{ only if the value of round constant rc(j) is 1}. $
	\end{enumerate}
	where $0 \leq j \leq (n/2)$. The first and the second step is represented by $\mathcal{R}_3$ and $\mathcal{R}_4$ in Figure~\ref{w2} respectively. \\
	After $k_2$ has been computed, we can compute each bits $k_3(j)$ applying the following steps: 
	\begin{enumerate}
	\item $ \cx(k_2(j-3)mod(n/2),k_1(j)), $
	\item $ \cx(k_2(j-4)mod(n/2),k_1(j)), $
	\item $ NOT(k_1(j)) \text{ only if the value of round constant rc(j) is 1}. $
	\end{enumerate}
	where $0 \leq j \leq (n/2)$. Similarly, we can proceed to compute the further round keys sequentially.
		
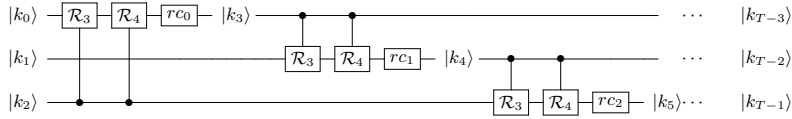
\begin{figure}[ht]
		\centering
			\scalebox{0.75}{
		$\Qcircuit @C=.8em @R=1.em {
			\lstick{\ket{k_0}} & \gate{ \mathcal{R}_3} & \gate{\mathcal{R}_4} & \gate{rc_0} & \rstick{\ket{k_3}} \qw& & & &\qw & \ctrl{1} & \ctrl{1} & \qw & \qw & \qw & \qw& \qw & \qw & \qw & \qw & \qw & \qw & \rstick{\cdots}& & &&\rstick{\ket{k_{T-3}}}\\
			\lstick{\ket{k_1}} & \qw & \qw &\qw &\qw & \qw & \qw &\qw &\qw &  \gate{ \mathcal{R}_3} & \gate{\mathcal{R}_4} & \gate{rc_1} & \rstick{\ket{k_4}} \qw & & & & \ctrl{1} & \ctrl{1} & \qw & \qw&\qw & \rstick{\cdots}& & &&\rstick{\ket{k_{T-2}}}\\
			\lstick{\ket{k_2}} & \ctrl{-2} & \ctrl{-2} & \qw & \qw & \qw & \qw &\qw &\qw & \qw & \qw & \qw & \qw & \qw & \qw & \qw & \gate{ \mathcal{R}_3} & \gate{\mathcal{R}_4} & \gate{rc_2} & \rstick{\ket{k_5}} \qw &  & \rstick{\cdots}& & &&\rstick{\ket{k_{T-1}}}
		}$}
		\caption{Circuit for key expansion with $3$ keywords. $rc_i$ represents the round dependent constants}
		\label{w3}
	\end{figure}

	For $m=3$ we have three key words $k_0,k_1,k_2$ each of size $n$. The round keys for further round can be computed as explained above for $m=2$ and the details are shown in Figure~\ref{w3}.

\begin{figure}[ht]
		\centering
			\scalebox{0.6}{
		$\Qcircuit @C=1em @R=1.em {
			\lstick{\ket{k_0}}&\gate{\mathcal{R}_1}&\gate{\mathcal{R}_3}&\gate{\mathcal{R}_4}&\gate{rc_0}&\rstick{\ket{k_4}}\qw&&&\qw&\ctrl{1}&\ctrl{1}&\qw&\qw&\qw&\qw&   \qw&\qw&\qw&\qw&\qw&\qw&\qw&\ctrl{3}&\qw&\qw&\qw&\qw&\qw&\qw&\rstick{\cdots}& & &&\rstick{\ket{k_{T-4}}}\\
			\lstick{\ket{k_1}}&\ctrl{-1}&\qw&\qw&\qw&\qw&\qw&\qw&\gate{\mathcal{R}_1}&\gate{\mathcal{R}_3}&\gate{\mathcal{R}_4}&\gate{rc_1}&\rstick{\ket{k_5}}\qw&&&\qw&\ctrl{1}&\ctrl{1}&\qw&\qw&\qw&\qw&\qw&\qw&\qw&\qw&\qw&\qw&\qw&\rstick{\cdots}& & &&\rstick{\ket{k_{T-3}}}\\
			\lstick{\ket{k_2}}&\qw&\qw&\qw&\qw&\qw&\qw&\qw&\ctrl{-1}&\qw&\qw&\qw&\qw&\qw&\qw&\gate{\mathcal{R}_1}&\gate{\mathcal{R}_3}&\gate{\mathcal{R}_4}&\gate{rc_2}&\rstick{\ket{k_6}}\qw&&&\qw&\ctrl{1}&\ctrl{1}&\qw&\qw&\qw&\qw&\rstick{\cdots}& & &&\rstick{\ket{k_{T-2}}}\\
			\lstick{\ket{k_3}}&\qw&\ctrl{-3}&\ctrl{-3}&\qw&\qw&\qw&\qw&\qw&\qw&\qw&\qw&\qw&\qw&\qw&\ctrl{-1}&\qw&\qw&\qw&\qw&\qw&\qw&\gate{\mathcal{R}_1}&\gate{\mathcal{R}_3}&\gate{\mathcal{R}_4}&\gate{rc_3}&\rstick{\ket{k_7}}\qw&&&\rstick{\cdots}& & &&\rstick{\ket{k_{T-1}}}
		}$}
		\caption{Circuit for key expansion with $4$ keywords. $rc_i$ represents the round dependent constants}
		\label{w4}
	\end{figure}
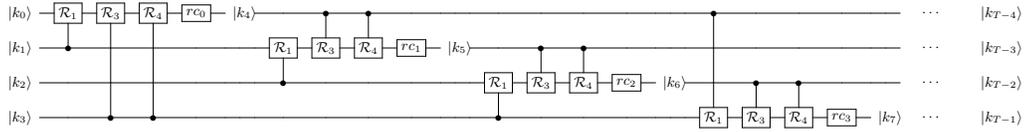

	For $m=4$ we have three key words $k_0,k_1,k_2,k_3$ each of size $n$. For the extra round keys we will require an extra step. $k_4(j)$ is computed applying the following steps:
	\begin{enumerate}
	\item $ \cx(k_1(j-1)mod(n/2),k_0(j)), $
	\item $ \cx(k_3(j-3)mod(n/2),k_0(j)), $
	\item $ \cx(k_3(j-4)mod(n/2),k_0(j)), $
	\item $ NOT(k_0(j)) \text{ only if the value of round constant rc(j) is 1} $
	\end{enumerate}
	where $0 \leq j \leq (n/2)$. The first step is the extra step required for $m=4$. The circuit is described in Figure~\ref{w4}.\\
	
	It can be easily calculated that for $ 1 $ round of key expansion we need  $mn$ \cx gates and $n'$ NOT gates (where $0\leq n' \leq n$ depending on the number of $1$'s in the round constant). In the complete implementation of \sio,  $(T-m)$  round-keys are generated as the first $m$ key words are used as first $m$ round keys. So, for $(T-m)$ rounds of key expansion we need $(T-m)(mn)$ \cx gates and $(T-m)n'$ $NOT$ gates.  
	
\begin{figure}[ht]
			\includegraphics[scale = 0.17]{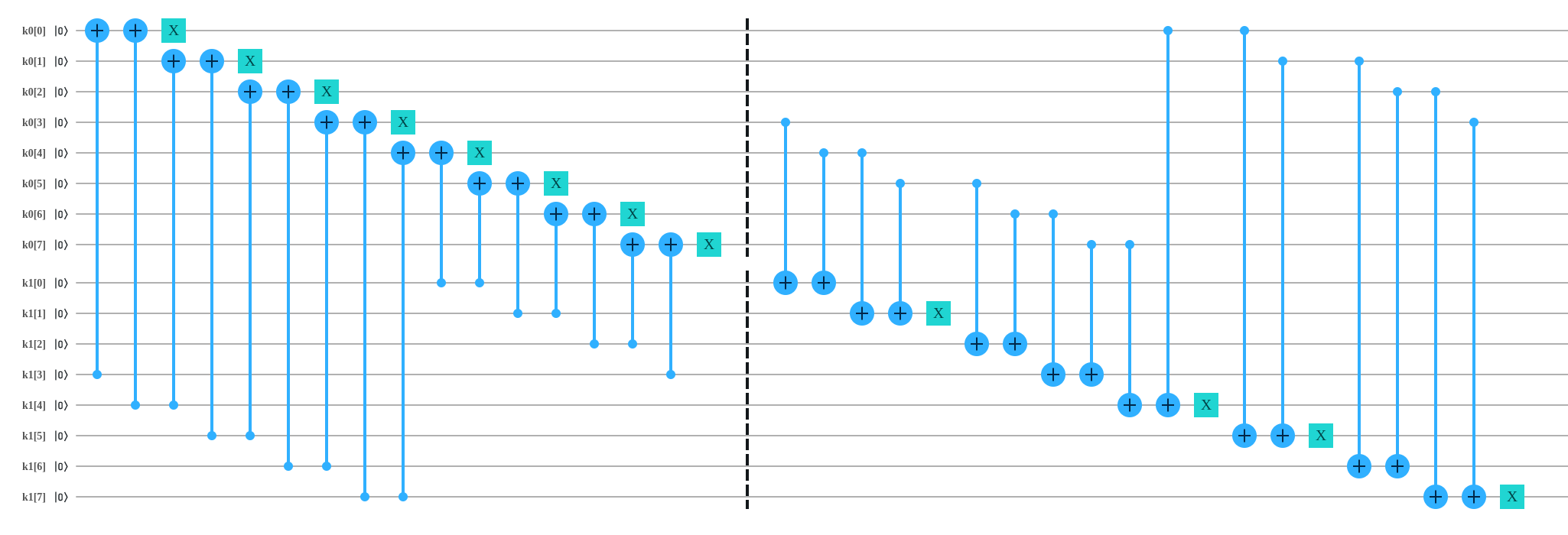}
			\caption{Key expansion for $m=2$ and key size $16$ split into two round keys of size $8$ each.}
			\label{keytri}
		\end{figure}

Consider the circuit in Figure~\ref{keytri}. This circuit represents reduced version of key expansion of \sio with key words $2$, as defined in Eqn.~\ref{keyexp}. The two round constants are assumed to be $c_0 = 11111111$ and $c_1 = 01001101$, which are implemented in the circuit by using $NOT $ gates. Barrier separates $k_3$ from $k_4$.

	\subsection{Circuit for \sio}
	\label{Scir}
	We implement \sio as a reversible circuit as reversibility is necessary for the cipher to be useful as a subroutine in Grover search.  Using the circuits developed for round function and key expansion we can now construct the circuit for full round \sio. 
	
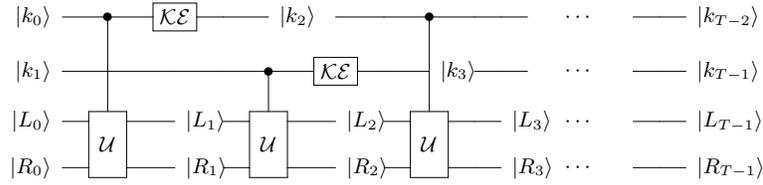
\begin{figure}[ht]
		\centering
		\scalebox{0.9}{
	$\Qcircuit @C=1.2em @R=1.2em {
		\lstick{\ket{k_0}}&\ctrl{2}&\gate{\mathcal{KE}}&\qw&\rstick{\ket{k_2}}\qw&&\qw&\ctrl{2}&\qw&\qw&\qw& \rstick{\cdots}&&&&\qw&\rstick{\ket{k_{T-2}}}\qw\\
		\lstick{\ket{k_1}}&\qw&\qw&\qw&\ctrl{1}&\gate{\mathcal{KE}}&\qw&\rstick{\ket{k_3}}\qw&&\qw&\qw&\rstick{\cdots}&&&&\qw&\rstick{\ket{k_{T-1}}}\qw\\
		\lstick{\ket{L_0}}&\multigate{1}{\mathcal{U}}&\rstick{\ket{L_1}}\qw&&\multigate{1}{\mathcal{U}}&\rstick{\ket{L_2}}\qw&&\multigate{1}{\mathcal{U}}&\qw&\rstick{\ket{L_3}}\qw&&\rstick{\cdots}&&&&\qw&\rstick{\ket{L_{T-1}}}\qw \\
		\lstick{\ket{R_0}}&\ghost{\mathcal{U}}&\rstick{\ket{R_1}}\qw&&\ghost{\mathcal{U}}&\rstick{\ket{R_2}}\qw&&\ghost{\mathcal{U}}&\qw&\rstick{\ket{R_3}}\qw&&\rstick{\cdots}&&&&\qw&\rstick{\ket{R_{T-1}}}\qw \\
	}$}
\caption{The circuit for implementing \sio with two key words. .}
\label{sicir}
	\end{figure}

	The input to the circuit is the key $K$ and the plaintext split into two halves $L_0,R_0$. The output of the circuit is the ciphertext $L_{T-1}, R_{T-1}$, where $T$ is the number of rounds. The size of $K,L, R$ are $mn,n,n$ respectively, where $m$ is either $2$ or $3$ or $4$. In Figure~\ref{sicir}, we draw the circuit considering $m=2$. Similar construction can be made for variants with $m=3$ and $m=4$. 	$\mathcal{U}$, is the round update function which consists of the three subroutines $\mathcal{F},\mathcal{G},\mathcal{H}$ and $\mathcal{KE}$ is the key expansion routine described in~\ref{keyex}.
	
\begin{figure}
				\centering
			\includegraphics[scale = .11]{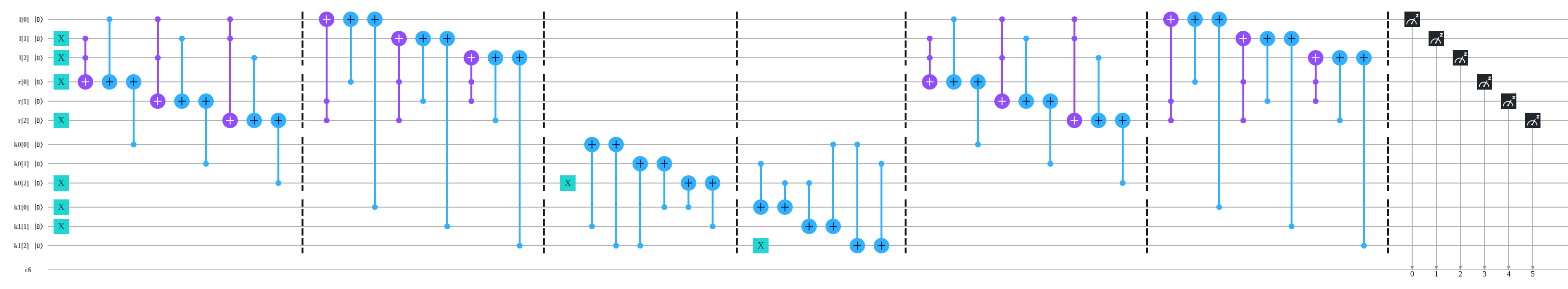}
			\caption{Circuit for $4$ rounds reduced \sio  with two key words} 
			\label{simcom}
	
		\vspace*{\floatsep}
		\includegraphics[scale = .3]{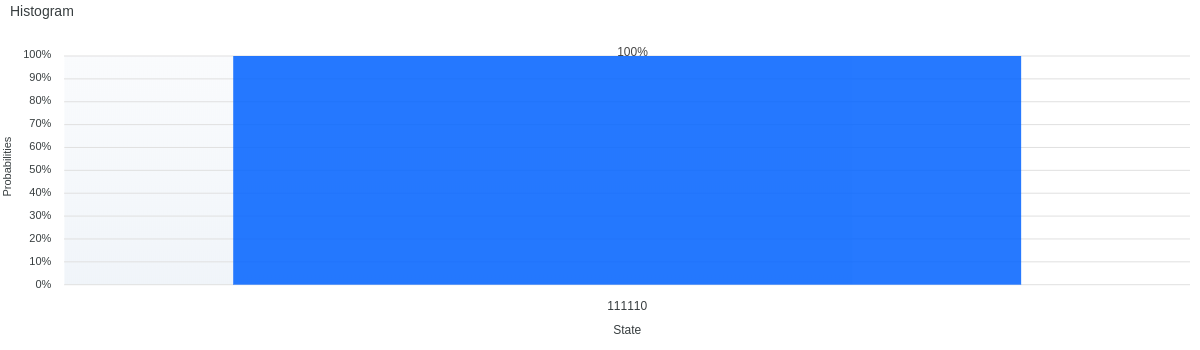}
		\caption{Measurement of circuit in Figure~\ref{simcom} in IBMQ Simulator. As expected the output after $ 4 $ rounds is $[011111]$.}
		\label{histo}
	\end{figure}

Figure~\ref{simcom} gives an implementation of the reduced version of \sio, with two key words. First, we run the circuit in IBMQ Simulator~\cite{ibmq}. We consider the key size and state size to be $6$ and the number of rounds to be $4$. The state update function is defined as 
$((L_{j+1},R_{j+1}) = (R_j\oplus (S^1(L_j) \& S^2(L_j)) \oplus S^0(L_j) \oplus k_j), L_j)$ and the key expansion is defined as $k_{j+2} = c_j \oplus  k_j  \oplus  S^{-1}(k_{j+1})  \oplus S^{-2}(k_{j+1})$ where $c_j$ are round dependent constants $[0,0,1]$	and $[0,0,1]$ for the third and fourth round respectively. Let the plaintext be $L_0 = [0,1,1], R_0 = [1,0,1]$ and the key words be $k_0 = [0,0,1], k_1 = [1,1,0]$. Then after four rounds the ciphertext will be $L_4 = [0,1,1], R_4 = [1,1,1]$. In IBMQ circuit, one should read the cipher text as $L_4=[L_4(0),L_4(1),L_4(2)]$ and $R_4=[R_4(0),R_4(1),R_4(2)]$. 
In Figure~\ref{histo} the output is shown as $[R_4(2),R_4(1),R_4(0),L_4(2),L_4(1),L_4(0)]$.
	
\begin{figure}
	\centering
	\includegraphics[scale = .25]{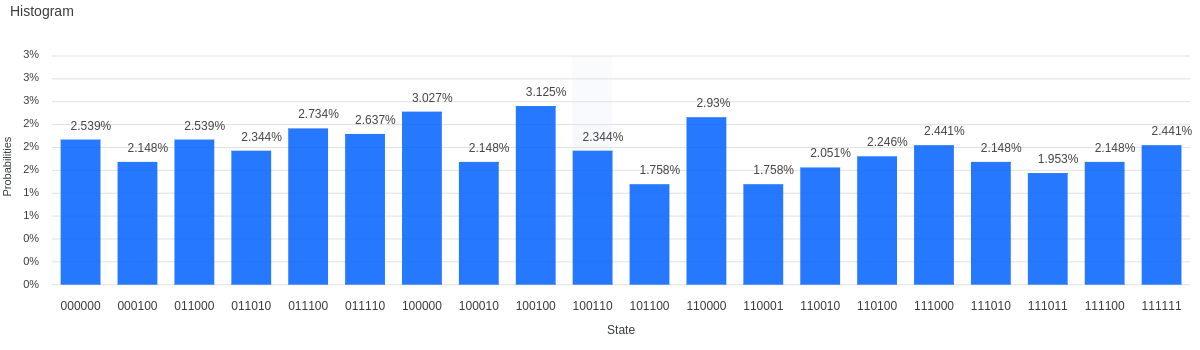}
	\caption{Measurement of circuit in Figure~\ref{simcom} in ibmq\_melbourne, the $14$ qubits actual processor.}
	\label{histo11}
	
\end{figure}

Then, we run the circuit (Figure~\ref{simcom}) in ibmq\_melbourne, the $14$ qubits actual processor. In contrary to the simulation (Figure~\ref{histo}), we observed a huge error in the result. Figure~\ref{histo11} shows the histogram we have obtained after running the circuit in ibmq\_melbourne for $ 1024 $ shots. The reason behind this  is the infidelity of the gates used in the circuit and the decoherence time of the qubits used in the actual processor, which has been pointed out in several works like~\cite{harper,tannu} and also in the IBMQ's official site.

	\subsection{Grover's oracle}
\begin{figure}[ht]
		\centering
		\scalebox{0.9}{
		$\Qcircuit @C=1.2em @R=1.2em {
			\lstick{\ket{K}}  &\qw&\qw&\qw&\qw&\ctrl{3}&\qw&\multigate{1}{\mathcal{SI}}&\qw&\qw&\qw&\qw&\qw&\multigate{1}{\mathcal{SI}^{\dagger}}&\qw&\qw&\ctrl{3}&\qw&\qw&\qw&\qw&\qw&\rstick{\ket{K}}\qw&\\
			\lstick{\ket{M_1}}&\qw&\qw&\qw&\qw&\qw&\qw&\ghost{\mathcal{SI}}&\qw&\qw&\qw&\ctrl{3}&\qw&\ghost{\mathcal{SI}^{\dagger}}&\qw&\qw&\qw&\qw&\qw&\qw&\qw&\qw&\rstick{\ket{M_1}}\qw&\\
			&&&&&&&&&\rstick{\ket{C_1}}&&&&&&&&&&&&&\\
			&&&&\lstick{\ket{0}}&\targ&\qw&\multigate{1}{\mathcal{SI}}&\qw&\qw&\qw&\qw&\qw&\multigate{1}{\mathcal{SI}^{\dagger}}&\qw&\qw&\targ&\qw&\rstick{\ket{0}}\qw&&&&\\
			\lstick{\ket{M_2}}&\qw&\qw&\qw&\qw&\qw&\qw&\ghost{\mathcal{SI}}&\qw&\qw&\qw&\ctrl{2}&\qw&\ghost{\mathcal{SI}^{\dagger}}&\qw&\qw&\qw&\qw&\qw&\qw&\qw&\qw&\rstick{\ket{M_2}}\qw&\\
			&&&&&&&&&\rstick{\ket{C_2}}&&&&&&&&&&&&&\\
			\lstick{\ket{-}}&\qw&\qw&\qw&\qw&\qw&\qw&\qw&\qw&\qw&\qw&\gate{=}&\qw&\qw&\qw&\qw&\qw&\qw&\qw&\qw&\qw&\qw&\qw&\rstick{\ket{-}}\gategroup{1}{4}{7}{21}{1.6em}{--}&\\
		}$}
		
		\caption{Grover's oracle for \sio using two plaintext-ciphertext pairs. Here, $ \mathcal{SI} $ represents the \sio encryption as described in~\ref{Scir}. The $(=)$ operator compares the output of the $\mathcal{SI} $ with the given ciphertexts and flips the target qubit if they are equal.}
		\label{twoinst}
	\end{figure}
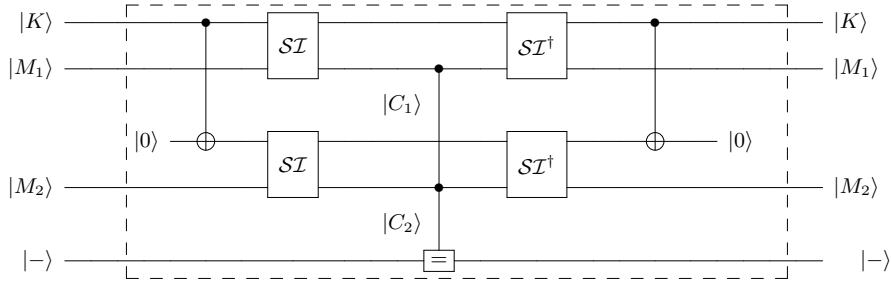

	In this subsection, we will discuss the implementation of Grover search on a block cipher under known plain text attack. Let $r$ pairs of plaintext and ciphertext be sufficient to successfully extract the unique solution. In this regard, we have to design an oracle that encrypts the given $r$ plaintexts under the same key and then computes a Boolean value which determines if all the resulting ciphertexts are equal to the given classical ciphertexts. This can be done by running the block cipher circuit $r$ times in parallel. Then the resultant ciphertexts  are compared with the classical ciphertexts. The target qubit will be flipped if the ciphertexts match. This is called Grover oracle. In Figure~\ref{twoinst}, the construction of such an oracle is given for two instances of plaintext-ciphertext pairs considering \sio block cipher. \\

	We implement the above idea, i.e.,  Grover oracle for our reduced version of \sio (Figure~\ref{simcom}) in the IBMQ simulator. We have not run the circuit for Grover in ibmq\_melbourne, as the output of the reduced \sio in this processor has already been masked with a huge amount of error (Figure~\ref{histo11}). 
	
\begin{figure}
		\centering
		\begin{subfigure}{.5\textwidth}
			\centering
			\includegraphics[scale = .35]{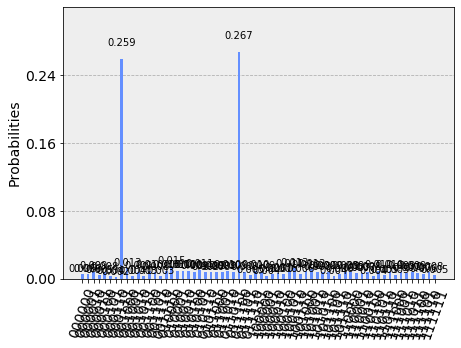}
			\caption{}
			\label{histo2}
		\end{subfigure}%
		\begin{subfigure}{.5\textwidth}
			\centering
			\includegraphics[scale = .35]{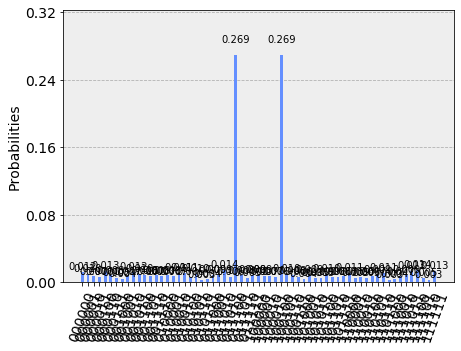}
			\caption{}
			\label{histo3}
		\end{subfigure}
		\caption{Histogram obtained after running Grover's on the reduced \sio described in Figure~\ref{simcom}.}
		
	\end{figure}

	Now, in this case, the plaintext is $M= [0,1,1,1,0,1]$ and the key is $K = [0,0,1,1,1,0]$. Then after four rounds the ciphertext will be $C= [0,1,1,1,1,1]$. In theory, this key $K = [0,0,1,1,1,0]$ will be obtained as the output of  Grover oracle. Figure~\ref{histo2} shows the outcome of  Grover applied to this circuit. Note that in the histogram two peaks appear; one for the exact key $K$ and another for an unknown key $K'=[1,1,1,0,0,0]$. This is because, the unknown key $K'$ also encrypts the plain text $M$ onto the cipher text $C$. To find the unique key, we use another plaintext-ciphertext pair, $M_1 = [0,0,1,1,0,1], C_1 = [1,1,0,0,1,1]$ under the same key. Figure~\ref{histo3} shows the outcome of  Grovers applied to this circuit. In this histogram also we got two peaks. Similar to~\ref{histo2}, one peak is for the exact key $K$ and another is for an unknown key $K'' = [0,0,1,0,0,1]$ as $K''$ encrypts $M_1$ to $C_1$ too. Note that the key common to both the pairs is $K = [0,0,1,1,1,0]$ which is the exact key. The full implementation is described in~\cite{grover1}. The functions Grover\_oracle and Grover\_Diffusion used in our code is taken from~\cite{daniel} by importing the file Our\_Qiskit\_Functions.

	\subsection{Resource Estimation}

	\subsubsection{Cost of implementing \sio}
	
	We first estimate the cost of implementing \sio as a circuit including the key expansion as well as the round function. As discussed above the round constants are implemented to the key expansion function using an adequate number of $NOT$ gates. We have not included the number of $NOT$ required to initialize the plaintext in our estimates as it depends on the given plaintext. Table~\ref{rcost} gives the cost estimates of implementing all \sio variants.\\
	\begin{table}[h]
		\centering
		\begin{tabular}{|c|c|c|c|c|c|}
			\hline
			\sio$2n/mn$ & \# $NOT$ & \# \cx & \# \tf & \# qubits& depth \\
			\hline
			\sio$ 32/64 $ & $ 448 $ & $ 2816$ & $ 512 $ & $96$ & $ 946 $\\
			\hline
			\sio$ 48/72 $ & $ 792 $ & $ 3312$ & $ 864 $ & $120$ &$ 1062 $\\
			\hline
			\sio$ 48/96 $ & $ 768 $ & $ 4800$ & $ 864 $ & $144$ &$ 1597 $\\
			\hline
			\sio$ 64/96 $ & $ 1248 $ & $ 5184$ & $ 1344 $ & $160$ & $ 1674 $\\
			\hline
			\sio$ 64/128 $ & $ 1216 $ & $ 7396$ & $ 1408 $ & $192$& $ 2643 $\\
			\hline
			\sio$ 96/96 $ & $ 2400 $ & $ 9792$ & $ 2496 $ & $192$& $ 4785 $\\
			\hline
			\sio$ 96/144 $ & $ 2448 $ & $ 10080$ & $ 2592 $ & $240$ & $ 3282 $\\
			\hline
			\sio$ 128/128 $ & $ 4224 $ & $ 17152$ & $ 4352 $ & $256$ &$ 8427 $\\
			\hline
			\sio$ 128/192 $ & $ 4224 $ & $ 17472$ & $ 4416 $ & $320$ &$ 5656 $\\
			\hline
			\sio$ 128/256 $ & $ 4352 $ & $ 26624$ & $ 4608 $ & $384$ & $ 8848 $\\
			\hline
		\end{tabular}
	\caption{Cost of implementing \sio variants}
	\label{rcost}
	\end{table}

\subsubsection{Cost of Grover oracle }
	
	Following \cite{aes3} we assume that $r = \ceil{k/(2n)}$ known plaintext-ciphertext pairs are sufficient to give us a unique solution, where $2n$ is the block size and $k=mn$ is the key size of the cipher. Here, one should mention that if $r = \ceil{k/(2n)}$ is an integer, then consider its next integer.  Grover's oracle consists of comparing the $2n$-bit outputs of the \sio instances with the given $r$ ciphertexts. This can be done using a $(2n\cdot r)$-controlled \cx gates (we neglect some $NOT$ gates which depend on the given ciphertexts). Following \cite{nathan}, we estimate the number of $ T $ gates required to implement a $t$-fold controlled $NOT$ gates as $(32 \cdot t - 84)$. 
	
	We use the decomposition of \tf gates to $7$ $T$-gates plus $8$ Clifford gates, a $T$-depth of $ 4 $ and total depth of $ 8 $ as in \cite{amy}. To estimate the full depth and the $T$-depth we only consider the depths of the \sio instances ignoring the multi controlled $NOT$ gate used in comparing the ciphertexts. We also need $(2\cdot (r-1) \cdot k)$ \cx gates to make the input key available to all the \sio instances in the oracle. The total number of Clifford gates is the sum of the Clifford gates used in the implementation of \sio and the $(2\cdot (r-1) \cdot k)$ \cx gates needed for input key. The cost estimates for all \sio variants are presented in Table~\ref{oraclecost}\\
	
	\begin{table}[ht]
		\centering
		\scalebox{0.85}{
		\begin{tabular}{|c|c|c|c|c|c|c|}
			\hline
			\sio$2n/k$ & $r$ & \# Clifford gates & \# $ T $ gates & $ T $-depth & full depth & \# qubits \\
			\hline
			\sio$32/64$ & $3$ & $ 19840 $ & $ 24492 $ & $ 12288 $ & $ 27180 $ & $161$\\
			\hline
			\sio$48/72$ & $2$ & $ 16560 $ & $ 27180 $ & $ 13824 $ & $ 28440 $ & $169$\\
			\hline
			\sio$48/96$ & $3$ & $ 33792 $ & $ 40812 $ & $ 20736 $ & $ 45860 $ & $241$\\
			\hline
			\sio$64/96$ & $2$ & $ 25620 $ & $ 41644 $ & $ 21504 $ & $ 44988 $ & $224$\\
			\hline
			\sio$64/128$ & $3$ & $ 52184 $ & $ 65196 $ & $ 33792 $ & $ 74994 $ & $321$\\
			\hline
			\sio$96/96$ & $2$ & $ 48768 $ & $ 75948 $ & $ 39936 $ & $ 89028 $ & $289$\\
			\hline
			\sio$96/144$ & $2$ & $ 50400 $ & $ 78636 $ & $ 41472 $ & $ 86104 $ & $337$\\
			\hline
			\sio$128/128$ & $2$ & $ 85760 $ & $ 129964 $ & $ 69632 $ & $ 151564 $ & $385$\\
			\hline
			\sio$128/192$ & $2$ & $ 87168 $ & $ 131756 $ & $ 70656 $ & $ 146272 $ & $449$\\
			\hline
			\sio$128/256$ & $3$ & $ 186880 $ & $ 205740 $ & $ 110592 $ & $ 246624 $ & $641$\\
			\hline
		\end{tabular}
		}
	\caption{Cost of Grover oracle}
	\label{oraclecost}
	\end{table}
\subsubsection{Cost of exhaustive key search }

Using the estimates in Table~\ref{oraclecost} of  Grovers oracle for the various variants, we provide the cost estimates for the full exhaustive key search on all variants in Table~\ref{exh}. We consider $\floor{\frac{\pi}{4}2^{k/2}}$ iterations of  Grovers operator. As in~\cite{aes2} we do not consider the depth of implementing the two multi-controlled $NOT$ gates while calculating the $T$-depth and overall depth. \\
\begin{table}[h]
	\centering
	\scalebox{0.83}{
	\begin{tabular}{|c|c|c|c|c|c|c|}
		\hline
		\sio$2n/k$ &  \# Clifford gates & \# $ T $ gates & $ T $-depth & Full depth & \# qubits \\
		\hline
		\sio$32/64$ &  $ 1.35 \cdot 2^{45.5} $ & $ 1.27 \cdot 2^{46} $ & $ 1.18 \cdot 2^{45} $ & $ 1.05 \cdot 2^{46.3} $ & $161$\\
		\hline
		\sio$48/72$ &  $ 1.01 \cdot 2^{49.65} $ & $ 1.03 \cdot 2^{50.45} $ & $ 1.01 \cdot 2^{49.4} $ & $ 1.05 \cdot 2^{50.37} $ & $169$\\
		\hline
		\sio$48/96$ &  $ 1.02 \cdot 2^{62.66} $ & $ 1.02 \cdot 2^{63.05} $ & $ 1.01 \cdot 2^{61.97} $ & $ 1.02 \cdot 2^{63.11} $ & $241$\\
		\hline
		\sio$64/96$ &  $ 1.02 \cdot 2^{62.27} $ & $ 1.01 \cdot 2^{63.08} $ & $ 1.10 \cdot 2^{61.9} $ & $ 1.07 \cdot 2^{63} $ & $224$\\
		\hline
		\sio$64/128$ &  $ 1.03 \cdot 2^{79.27} $ & $ 1.02 \cdot 2^{79.7} $ & $ 1.06 \cdot 2^{78.6} $ & $ 1.03 \cdot 2^{79.8} $ & $321$\\
		\hline
		\sio$96/96$ &  $ 1.02 \cdot 2^{63.2} $ & $ 1.04 \cdot 2^{63.85} $ & $ 1.02 \cdot 2^{62.9} $ & $ 1.02 \cdot 2^{64} $ & $289$\\
		\hline
		\sio$96/144$ &  $ 1.05 \cdot 2^{87.2} $ & $ 1.06 \cdot 2^{87.9} $ & $ 1.22 \cdot 2^{86.7} $ & $ 1.03 \cdot 2^{88} $ & $337$\\
		\hline
		\sio$128/128$ &  $ 1.03 \cdot 2^{80} $ & $ 1.14 \cdot 2^{80.5} $ & $ 1.17 \cdot 2^{79.51} $ & $ 1.12 \cdot 2^{80.7} $ & $385$\\
		\hline
		\sio$128/192$ &  $ 1.04 \cdot 2^{112} $ & $ 1.17 \cdot 2^{112.5} $ & $ 1.19 \cdot 2^{111.51} $ & $ 1.08 \cdot 2^{112.7} $ & $449$\\
		\hline
		\sio$128/256$ &  $ 1.05 \cdot 2^{145.1} $ & $ 1.11 \cdot 2^{145.2} $ & $ 1.07 \cdot 2^{144.3} $ & $ 1.12 \cdot 2^{145.4} $ & $641$\\
		\hline
	\end{tabular}
}
\caption{Cost estimates of Grovers algorithm with $\floor{\frac{\pi}{4}2^{k/2}}$ oracle iterations with a success probability negligibly close to $1$.}
\label{exh}
\end{table}

	\section{Conclusion}
	\label{conclusion}
	In this work we presented an implementation of Grovers search algorithm on \sio. We first provided a reversible circuit of all variants of \sio. Then these circuits were used to estimate the cost of attacking \sio using  Grover's algorithm. The plausible values the overall circuit depth suggested by NIST~\cite{nist} is between $2^{40}$ and $2^{96}$ and we assume that this depth is an upper bound on the total depth of a quantum attack. The overall circuit depth of all the implementations presented in this work except for $\sio128/192$ and $\sio128/256$ lie in this range. This work is aimed at using the minimal number of qubits, in future it would be interesting to re-estimate the cost by reducing the depth of the implementations at the expense of some extra qubits by using other choices of decomposition of \tf gates, e.g., as in \cite{peter} which has a $T$-depth of $1$ and an overall depth of $7$ but uses $18$ Clifford gates and $4$ ancilla qubits.

\end{document}